\documentclass[
twocolumn,
amsmath,amssymb,
aps,pra 
]{revtex4-2}

\usepackage{graphicx}
\usepackage{dcolumn}
\usepackage{bm}
\usepackage[dvipsnames]{xcolor}
\usepackage{graphicx}
\usepackage{float} 
\usepackage[none]{hyphenat}
\usepackage{lipsum}
\usepackage{float}
\usepackage{xcolor}
\usepackage{cancel}
\usepackage[colorlinks=true, 
linkcolor=blue,          
citecolor=blue,        
filecolor=blue,      
urlcolor=blue]{hyperref}

\usepackage[all]{hypcap}

\begin{document}
	
	\preprint{APS/123-QED}
	
	\title{Generalized Time-Varying Drude Model for Dispersive and Lossy Modulations}

	\author{A. Ganfornina-Andrades}
	\author{J. E. Vázquez-Lozano}
	\author{I. Liberal}
		\author{S. A. R. Horsley.}
	\affiliation{Department of Electrical, Electronic and Communications Engineering,\\
		Institute of Smart Cities (ISC), Public University of Navarre (UPNA), 31006 Pamplona, Spain.\\
	Department of Physics and Astronomy, University of Exeter, Stocker Road, Exeter EX4 4QL, United Kingdom.}
	
	\begin{abstract}
We develop a generalization of the time-varying Drude model, treating carrier density, effective mass, and collision rate as explicit functions of time. We derive expressions for polarization, susceptibility, displacement, and permittivity in different domains. Our analysis reveals that non-adiabatic modulations and time-dependent losses induce rich and distinct behaviors, leading to temporal blurring, selective gating and suppression, and low-frequency spectral reshaping. Besides underpinning and upgrading the current framework on photonics of time-varying media, this model may be useful in the design and fitting theoretical models with experimental realizations.
	\end{abstract}

	\maketitle
	\sloppy
	
\section{INTRODUCTION}
Recent nanophotonics research has explored ultra-fast modulations of the macroscopic material responses, leading to the so-called fields of time-varying media \cite{galiffi2022photonics}, temporal metamaterials \cite{yuan2022temporal}, 4D optics \cite{engheta2020metamaterials,engheta2023four} and/or spacetime metamaterials \cite{caloz2019spacetime1,caloz2019spacetime2}. In the non-adiabatic regime, temporal modulations enable generalized wave manipulations, simultaneously combining beam-shaping, amplification, frequency shifting and pulse shaping \cite{galiffi2019broadband,taravati2022microwave,moreno2024space,bahrami2025arbitrary}. Similarly, breaking temporal symmetries fundamentally changes the context of light-matter interactions, resulting in modified spatio-temporal symmetries and conserved quantities  \cite{pendry2021gain,pendry2022photon,ortega2023tutorial,liberal2024spatiotemporal,jajin2024symmetries}. Moreover, ultra-fast modulations of the optical parameters expand the possibilities of quantum state engineering \cite{pendry2022photon,lyubarov2022amplified,dikopoltsev2022light,vazquez2023shaping,liberal2023quantum,ganfornina2024quantum,sustaeta2025quantum} and thermal emission \cite{vazquez2023incandescent,vertiz2025dispersion,liberal2025can}.


Crucially, dispersion—a quality inherent to optical media—becomes relevant in temporal metamaterials—a feature already recognized in early studies \cite{fante1973propagation,yablonovitch1973spectral,yablonovitch1974self,jiang1975wave,stepanov1976dielectric}. This role is specially pronounced in epsilon-near-zero (ENZ) materials \cite{liberal2017near,kinsey2019near,reshef2019nonlinear,lobet2020fundamental}, where the strong and ultra-fast nonlinear effects facilitate temporal modulations.

Among ENZ platforms, transparent conducting oxides (TCOs) \cite{alam2016large,zhou2020broadband,bohn2021spatiotemporal,harwood2025space,xu2025high,bohn2021all,tirole2022saturable,tirole2023double,liu2021photon,ball2025space,jaffray2025spatio} have emerged as leading candidates for several experimental demonstrations. These include temporal \cite{alam2016large,zhou2020broadband} and spatio-temporal \cite{bohn2021spatiotemporal,xu2025high} refraction, diffraction from syntehtic motion \cite{harwood2025space}, plasmon frequency shifts \cite{bohn2021all}, spectral manipulation with saturable time-varying mirrors \cite{tirole2022saturable}, temporal analogs of the double-slit experiment \cite{tirole2023double}, photon acceleration \cite{liu2021photon}, space–temporal knives \cite{ball2025space}, and spatio-spectral fission \cite{jaffray2025spatio}. These results emphasize the need for accurate theoretical models, since ENZ media are inherently dispersive and lossy \cite{javani2016real}, and the role of dispersion and loss under temporal modulation must therefore be properly accounted for \cite{torrent2020strong,hayran2022omega,koutserimpas2024time,sloan2024optical}.

The most widely adopted theoretical framework relies on a time-modulated Drude model, in which an explicit temporal dependence is introduced in the plasma frequency \cite{horsley2023eigenpulses,allard2025broadband,hooper2025quasi,feinberg2025plasmonic,verde2025optical,sustaeta2025near}. This approach has enabled several theoretical advances, including eigenpulses in time-varying media \cite{horsley2023eigenpulses}, dispersive \cite{allard2025broadband,hooper2025quasi} and plasmonic time crystals \cite{feinberg2025plasmonic}, time-varying plasmonic resonances \cite{verde2025optical}, and recent extensions of macroscopic quantum electrodynamics (mQED) \cite{horsley2025macroscopic}. Temporal modulation of alternative dispersion models, such as Lorentz-type responses, has also been explored in the context of Kramers–Kronig relations \cite{solis2021time,solis2021functional}, lattice resonances \cite{de2025lattice}, and unconventional frequency generation at natural resonances \cite{rizza2024harnessing}. However, time-varying Drude models based on an explicit modulation of the plasma frequency presents its own implicit assumptions and approximations. First, the plasma-frequency modulation speed is limited because higher-order \cite{horsley2025macroscopic} time derivatives are usually truncated. Second, temporal modulation of other parameters, such as the collision rate, is often neglected. However, recent studies highlight its key role in the nonlinear response of TCOs \cite{un2025optical}.

In this work, we generalize the time-varying Drude model to include arbitrary temporal modulation of all physical parameters involved, thus removing implicit assumptions. We derive closed-form expressions for the permittivity in time-frequency, two-times, and two-frequencies domains. Our results highlight new phenomena arising from both non-adiabatic effects and time-dependent losses, with implications in both the frequency domain and transient responses. We conclude with a study of wave reflection from a thin metallic slab with time-varying permittivity.

\section{TIME-VARYING DRUDE MODEL}

\begin{figure}[h!]
	\includegraphics[scale=0.5]{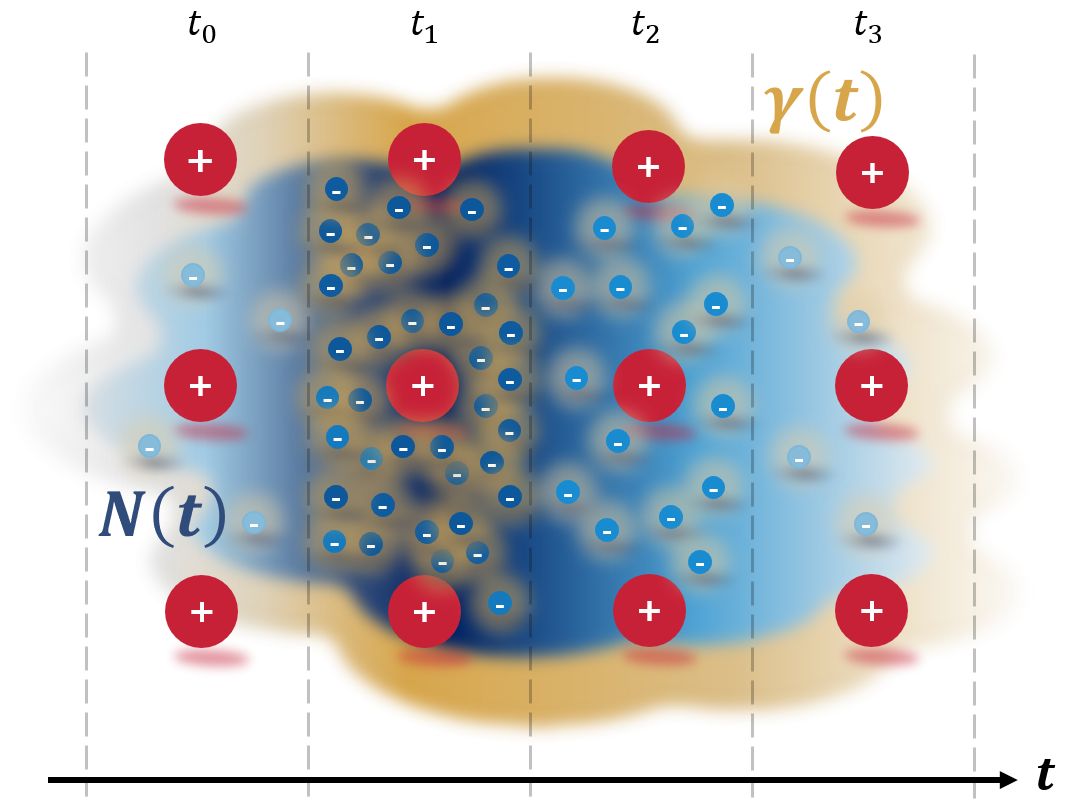}
	\caption{\label{tim_varying_Drude_model}Conceptual picture of the time evolution of a metallic system across four distinct frames $t_0<t_1<t_2<t_3$. In each frame, the carrier population $N(t)$ and collision rate $\gamma(t)$ are modulated differently. Atomic centers (ions) are depicted as red spheres, carriers (electrons) as smaller blue particles. The varieties of blue shading represents the electron density, whereas the shades of yellow depicts the collision rate, with darker regions corresponding to higher populations and collision rates, respectively. Temporal changes in $\gamma(t)$ and $N(t)$ influence the overall material response.}
\end{figure}

From a theoretical perspective, metals are commonly described by the Drude model \cite{drude1900elektronentheorie}, where free electrons of effective mass $m^*$ are damped by a collision rate $\gamma$. In this work, we extend this description to dispersive media with time-varying properties. We show that the dynamics in time-varying metals—conceptually shown in Fig.~\ref{tim_varying_Drude_model} by the modulation of $N(t)$ and $\gamma(t)$—admit closed-form solutions, making the time-varying Drude model a convenient framework for studying ultra-fast modulation and memory effects in metals.

\subsection{Equation of motion}
We begin by revisiting the classical equation of motion for an electron subject to a time-dependent effective mass $m^*(t)$ associated with alterations in the band structure of metals, a time-varying damping coefficient $\gamma(t)$ which reflects the collision rate, and an external electric field $\mathbf{E}(t)$. The corresponding equation reads:
\begin{equation}
	\frac{d}{dt}\left[m^*(t)\mathbf{v}(t)\right] + m^*(t)\gamma(t)\mathbf{v}(t) = q\mathbf{E}(t),
		\label{equation_of_motion}
\end{equation}
where $\mathbf{v}(t)$ is the charge carrier velocity and $q$ the electron charge. Assuming that this time-varying Drude material is causal, by standard manipulation—dividing by $m^*(t)$ (assumed to be non-zero in any time instant), identifying the integrating factor, and solving the resulting first-order ODE—we obtain a non-Markovian expression for $\mathbf{v}(t)$ involving, not only the present time, but an integration over the electric field’s past history, weighted by a memory kernel $I_\gamma(t')$:
\begin{equation}
\mathbf{v}\left(t\right)=\frac{q}{m^*\left(t\right){{I_{\gamma}\left(t\right)}}}\int_{-\infty}^{t}{dt'}\,{{I_{\gamma}\left(t'\right)}}{\mathbf{E}\left(t'\right)},
		\label{velocity}
\end{equation}
where $I_\gamma(t) = \exp\left[\int^t_{-\infty} \gamma(u)du\right]$ is the integrating factor accounting for the cumulative damping up to time $t$.
\subsection{Polarization}
Taking the current density for a set of charges moving uniformly at velocity $\mathbf{v}\left(t\right)$, so that $\mathbf{j}(t) = qN(t)\mathbf{v}(t)$, and assuming each charge obeys the aforementioned equation of motion for unidirectional fields, the polarization $P(t)$ is obtained via $\partial_t P(t) = j(t)$. A double integration process; the first over the past evolution of the electric field, and the second accounting for the temporal modulation of the material parameters, yields to:
\begin{equation}
P\left(t\right)=\varepsilon_{0}\int_{-\infty}^{t}dt'\thinspace{\frac{\omega_{p}^{2}\left(t'\right)}{I_{\gamma}\left(t'\right)}}\int_{-\infty}^{t'}{dt''\thinspace {I_{\gamma}\left(t''\right)} E\left(t''\right)},
	\label{polarization_before_integrating}
\end{equation}
where $\omega_p^2(t) = [q^2 N(t)]/[\varepsilon_0 m^*(t)]$ represents the plasma frequency of the system. From this derivation of $\omega_p(t)$, it is noticed that it can be modulated either via the carrier density or the effective mass \cite{galiffi2025electrodynamics}. 

Notably, Equation~\eqref{polarization_before_integrating} has a nested, convolution-like structure: the inner integral captures the field memory for a time-dependent collision rate, while the outer integral incorporates the plasma-frequency modulation. In the limits of constant $\gamma(t)$, constant $\omega_p^2(t)$, or both, it reduces to standard exponential convolution, simple rescaling, or the linear Drude response, respectively.

\section{SUSCEPTIBILITY AND PERMITTIVITY}
In conventional settings where the optical properties of a medium are static, the theoretical description involves a single time variable, which can be straightforwardly related to a single-frequency representation. However, when the optical properties evolve dynamically, temporal modulations of the material parameters introduce an additional variable, fairly complicating the analysis. Consequently, several complementary formalisms may be employed for addressing the same physical phenomena. In the following sections, using Laplace formalism, we derive the polarization field, the susceptibility, displacement field, and permittivity of the time-varying Drude model in the time--frequency, two-times, and two-frequencies domains~(\ref{sec:mixed}, \ref{sec:two_times}, \ref{sec:two_frequencies}).

\subsection{Time-frequency domain}\label{sec:mixed}
The mixed time–frequency description $\varepsilon(t,\omega)$ serves as a bridge between the modulation of parameters and their macroscopic response. While $\varepsilon(\omega)$ characterizes stationary systems through their spectral dispersion, the presence of an explicit time dependence in $\omega_p(t)$ and $\gamma(t)$ requires an extension that accounts for both instantaneous and accumulated effects of modulation. Hence, in this domain, $t$ stands for the observation time, and $\omega$ indicates how different frequencies contribute.

Since the static Drude model has a pole at $\omega=~0$, we choose to work by expressing the electric field in terms of Laplace transform $E(t)=(2\pi i)^{-1}\int^{\xi+i\infty}_{\xi-i\infty}ds \tilde{E}(s)e^{st}$. Upon substituting this expression into Eq.~(\ref{polarization_before_integrating}), the exponential factor is rearranged, isolating the $t$ dependence, leading to:
\begin{equation}
	P\left(t\right)=\frac{\varepsilon_{0}}{2\pi i}\int^{\xi+i\infty}_{\xi-i\infty} ds\,E\left(s\right)\chi(t,s)e^{st},
	\label{polarization_mixed_laplace}
\end{equation}
where the mixed Laplace-time susceptibility kernel is:
\begin{equation}
	\chi\left(t,s\right)= \int_{-\infty}^{t}dt'\int_{-\infty}^{t'}{dt''\thinspace {\omega_{p}^{2}\left(t'\right)}\thinspace{\frac{I_{\gamma}\left(t''\right)}{I_{\gamma}\left(t'\right)}}e^{s(t''-t)}}.
	\label{kernel_mixed_laplace}
\end{equation}
Then, by defining the function $f(t,s)=\omega_{p}^{2}(t)I(t,s)$, with  $I(t,s)=~\int_{-\infty}^{t}dt'\thinspace\left[I_{\gamma}\left(t'\right)/I_{\gamma}\left(t\right) \right]\thinspace e^{-s\left(t'-t\right)}$, the kernel (\ref{kernel_mixed_laplace}) can be decomposed into a first-order term $\chi_{1}$, 
\begin{equation}
\chi_{1}\left(t,s\right)=\frac{1}{s}f\left(t,s\right),
	\label{chi_1_mixed}
\end{equation}
and a higher-order correction $\chi_{2}$:
\begin{equation}
\chi_{2}\left(t,s\right)=-\frac{1}{s}\int_{-\infty}^t dt'\frac{d}{dt'}\left[f(t',s)\right]e^{s(t'-t)}.
	\label{chi_2_mixed}
\end{equation}
Next, we perform a Laplace-to-Fourier substitution projecting $s$ onto the imaginary axis i.e., $s\mapsto-i\omega + 0^{+}$, where the infinitesimal \(0^{+}\) indicates the limit \(\xi \to 0^{+}\) and enforces the causal prescription (Re $(s)>0$). The mixed time–frequency kernel in (\ref{kernel_mixed_laplace}) then becomes
\begin{equation}
	\begin{gathered}
\chi\bigl(t,\omega\bigr)
= \frac{f(t,\omega)}{-i\omega+0^+}-\frac{1}{-i\omega+0^+}\\
\cdot
\int_{-\infty}^{t} dt' \frac{d}{dt'}\left[f(t',\omega)\right]e^{-i(\omega +0^+)\left(t'-t\right)}.
	\end{gathered}
	\label{kernel_mixed}
\end{equation}

As for the static case \cite{poon2009kramers}, Eq.~(\ref{kernel_mixed}) highlights that the pole at $\omega=0$ must be be correctly handled to recover a causal response. While the influence of this pole can be neglected when the analysis is restricted at optical frequencies, it must be taken into account to recover fully time-domain expressions and low frequency radiation. Noticing $\omega \mapsto \omega + i0^{+}$, the polarization field reads
\begin{equation}
	P(t) = \varepsilon_{0} \int_{-\infty}^{\infty} \frac{d\omega}{2\pi}
	\chi(t,\omega)E(\omega)e^{-i\omega t}.
	\label{polarization_mixed}
\end{equation}
The expressions in Eqs.~(\ref{kernel_mixed}) and (\ref{polarization_mixed}) provide the causal polarization response under a time-varying Drude model, formulated via a mixed time–frequency susceptibility kernel $\chi(t,\omega)$ that explicitly incorporates plasma frequency $\omega_{p}(t)$ and collision rate $\gamma(t)$ modulation. For a linear response \mbox{$D(t)=\varepsilon_{0}\varepsilon_{\infty}E\left(t\right)+P\left(t\right)$}:
\begin{equation}
D\left(t\right)=\varepsilon_{0}\thinspace\int_{-\infty}^{+\infty}\frac{d\omega}{2\pi}\varepsilon\left(t,\omega\right)E\left(\omega\right)e^{-i\omega t},
	\label{D}
\end{equation}
where the time-frequency permittivity equals to:
\begin{equation}
\varepsilon\left(t,\omega\right)=\varepsilon_{\infty}+\chi\left(t,\omega\right).
	\label{varepsilon}
\end{equation}
In this time-frequency picture, the permittivity~(\ref{varepsilon}) weights the spectral components of the external electric field $E(\omega)$ to yield the displacement field $D(t)$. Noticing $f(t,\omega)=\omega_{p}^{2}(t)I(t,\omega)$, from Eq.~(\ref{kernel_mixed}), the permittivity in equation (\ref{varepsilon}) can be rewritten as follows:
\begin{widetext}
	\begin{equation}
			\varepsilon\left(t,\omega\right)=\varepsilon_{\infty}+\frac{i}{\omega}\thinspace\left[\omega_{p}^{2}\left(t\right)I\left(t,\omega\right)\right]-\frac{i}{\omega}\int_{-\infty}^{t}dt'\thinspace\frac{d}{dt'}\left[\omega_{p}^{2}\left(t'\right)I\left(t',\omega\right)\right]e^{-i\omega\left(t'-t\right)},
			\label{epsilon_two_terms}
		\end{equation}
\end{widetext}
where $I(t,\omega)=\int_{-\infty}^{t}dt'\thinspace\left[I_{\gamma}\left(t'\right)/I_{\gamma}\left(t\right)  \right]\thinspace e^{-i\omega\left(t'-t\right)}$. 
	 In this theoretical framework, Eq.~(\ref{epsilon_two_terms}) describes a time-varying permittivity, in which all the relevant parameters behave as dynamic macroscopic quantities rather than static ones. A closer examination of Eq.~(\ref{epsilon_two_terms}) reveals that the first two terms on its right-hand side (r.h.s.) correspond to a non-local temporal, non-separable contribution due to $\gamma(t)$, while the third term accounts for what we shall hereafter refer to as higher-order corrections.
 
\setlength{\parskip}{0pt}
Without loss of generality, we consider a temporal modulation of the macroscopic parameters such that the plasma frequency squared, $\omega_p^2(t)$, follows a pulsed profile (see Appendix~\ref{sec:figures}). This modulation is defined by a reference value $\omega_0^2$, which, for the chosen background $\varepsilon_{\infty}$, corresponds to the ENZ frequency of the unmodulated medium, and by an amplitude $\delta\omega_{p,1}^2$. The profile can be made symmetric or asymmetric via two characteristic times: a loading time $\tau_L$, primarily determined by the medium's response to pump–probe excitation, and a decay time $\tau_D$, set by the intrinsic relaxation dynamics. The amplitude $\delta\omega_{p,1}^2$ is chosen within the limits reported in~\cite{bohn2021spatiotemporal,bohn2021all,horsley2023eigenpulses,horsley2025macroscopic}. Regarding the collision rate $\gamma(t)$, we assume a profile similar to $\omega_p(t)$ but roughly an order of magnitude smaller, consistent with lightly damped metals~\cite{bohn2021all}. This choice is physically justified as follows: although $\gamma$ is dominated by electron–phonon interactions and depends weakly on carrier density $N(t)$, it is strongly influenced by lattice temperature \cite{alabastri2017controlling}; optical pumping transiently heats the material, increasing $\gamma(t)$.

\begin{center}
	\begin{figure}[H]
		\centering
		\includegraphics[scale=0.94]{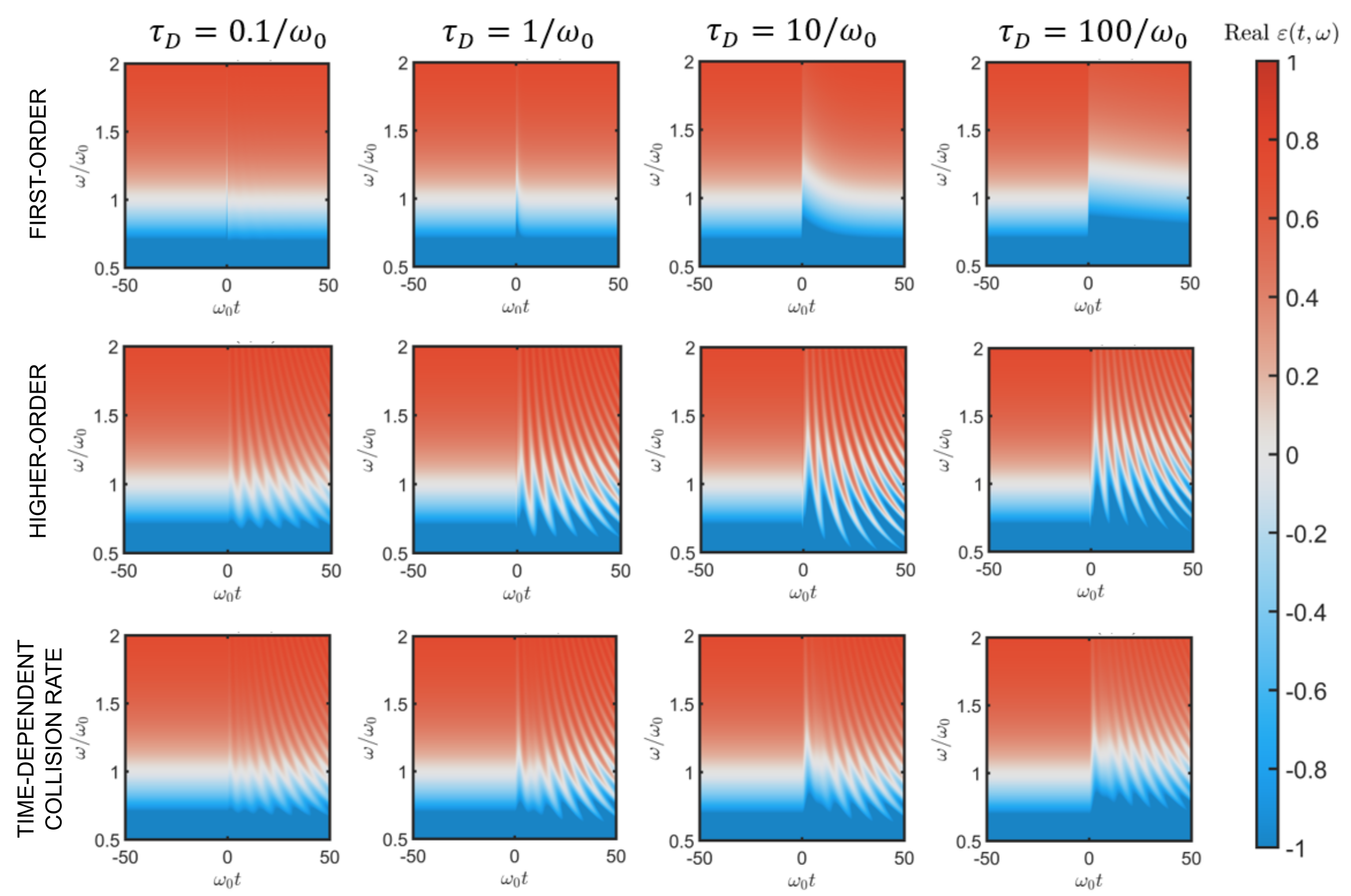}
		\caption{
			\label{three_models_epsilon_t_w}
			Real-part permittivity evolution across three time-varying Drude models with $\omega^2_{p}(t)=\omega^2_0+\delta\omega^2_{p,1}e^{-t/\tau_L}[1+\tanh(t/\tau_D)]$ so that $\omega^2_{p,\text{max}}=1.4\omega^2_0$. Upper row: first-order (Stepanov) model~\cite{stepanov1976dielectric}, assuming constant collision rate $\gamma=0.1\omega_0$ and time-varying plasma frequency $\omega^2_{p}(t)$. 
			Middle row: higher-order model~\cite{horsley2025macroscopic}, including non-instantaneous contributions arising from rapid temporal variations of $\omega_p^2(t)$ and constant collision rate $\gamma=0.1\omega_0$. 
			Bottom row: fully time-dependent model, where $\gamma(t)$ evolves dynamically according to $\gamma(t)=0.1\omega_p(t)$, with a slightly different modulation amplitude $\delta\omega^2_{p,2}$, so that $\gamma_{max}=0.14\omega_0$. This coupling gives rise to a non-separable kernel $I(t,\omega)$ and loss-induced temporal blurring. In all cases, $\omega_0\tau_L=0.1$. The four columns, from left to right, correspond respectively to $\omega_0\tau_D=0.1$, $1$, $10$, and $100$, illustrating the transition from ultra-short symmetric modulations to asymmetric profiles dominated by the hyperbolic-tangent envelope.
		}
	\end{figure}
\end{center}

A comparison between a first-order~\cite{stepanov1976dielectric}, a higher-order ~\cite{horsley2025macroscopic}, and time-dependent collision rate time-varying Drude model is shown in \hyperref[tim_varying_Drude_model]{Fig.~\ref*{three_models_epsilon_t_w}} showing the real part of $\varepsilon(t,\omega)$, so that we can track the dynamical evolution of ENZ region for the three models.


The first row corresponds to the Stepanov model~\cite{stepanov1976dielectric}, an early attempt to describe the optical response of a time-modulated plasma within the Drude framework, here referred to as a first-order model. 
	In this model, the collision rate $\gamma$ is held constant, while the carrier density—and consequently $\omega_p(t)$—is assumed to vary slowly in time compared to the optical oscillation period ($\dot{N}/N \ll \omega$). 
	By neglecting higher-order effects, the integral term in Eq.~(\ref{epsilon_two_terms}), which involves the temporal derivative of $\omega_p^2(t') I(t',\omega)$, is omitted. Consequently, the permittivity becomes a separable function, taking the conventional Drude-like form~\cite{bohn2021all,tirole2022saturable,tirole2023double,harwood2025space,jaffray2025spatio,horsley2023eigenpulses,allard2025broadband,hooper2025quasi,feinberg2025plasmonic,verde2025optical,sustaeta2025near}:
\begin{equation}
	\varepsilon(t,\omega) \approx \varepsilon_{\infty} - \frac{\omega_{p}^{2}(t)}{\omega(\omega+i\gamma)}.
\end{equation}

\onecolumngrid
\clearpage        
\twocolumngrid

Therefore, the first-order approximation provides a relatively adiabatic description of the time-varying permittivity, as shown in the first row of \hyperref[three_models_epsilon_t_w]{Fig.~\ref*{three_models_epsilon_t_w}}. 
	The modulation of $\omega_p^2(t)$ primarily results in a shift of the frequency where the real part of the permittivity crosses zero, moving toward higher values as time progresses. 
	For increasing decay times $\tau_D$, this shift becomes both smoother and more sustained, reflecting the slower relaxation of the carrier density. 
	Consequently, $\mathrm{Re}\,\varepsilon(t,\omega)$ exhibits an extended near-zero region that evolves from a short, symmetric modulation ($\tau_L=\tau_D=0.1/\omega_0$) into a longer-lived, asymmetric profile dominated by the hyperbolic-tangent envelope ($\tau_D = 100/\omega_0$).

However, in the regime of ultra-short modulation times (as it is the case), the adiabatic approximation ceases to hold. 
	Neglecting the integral term in Eq.~(\ref{epsilon_two_terms}) eliminates the strongly non-linear behavior associated with rapid variations in $\omega_p^2(t)$, which can induce significant temporal dispersion and oscillatory features in the permittivity. 
	To address these effects, a higher-order formulation must be adopted~\cite{horsley2025macroscopic}. 
	In particular, for a constant collision rate $\gamma$ and rapidly varying $\omega_p^2(t)$, the time–frequency permittivity develops oscillations, consistent with inclusion of the full integral term in Eq.~(\ref{epsilon_two_terms}), yielding~\cite{horsley2025macroscopic}:
\begin{equation}
	\varepsilon(t,\omega) = \varepsilon_{\infty} - \frac{\omega_{p}^{2}(t)}{\omega(\omega+i\gamma)}+\frac{\int_{-\infty}^{t}dt'\thinspace \dot{\omega}_p^2\left(t'\right)e^{-i\omega\left(t'-t\right)}}{\omega(\omega+i\gamma)}.
\end{equation}
This higher-order model evidences the non-instantaneous dynamics arising from an instantaneous modulation of the plasma frequency $\omega_{p}(t)$ and it is shown in the second row of \hyperref[three_models_epsilon_t_w]{Fig.~\ref*{three_models_epsilon_t_w}}, 
	so that leads to oscillations in the time–frequency permittivity. 
	Contrary to intuitive expectations, the most abrupt modulations do not yield the strongest oscillations, as evident from the ultra-short and symmetric case. 
	Instead, for non-trivial values of $\tau_D$, the shift in the zero of $\mathrm{Re}\,\varepsilon(t,\omega)$ spans a wider frequency range—particularly toward lower frequencies—relative to $\omega_0$. 
	For longer decay times $\tau_D$, these oscillations persist over time, as the hyperbolic-tangent envelope becomes dominant in shaping the profile of $\omega_p^2(t)$.

Finally, when both the plasma frequency $\omega_{p}(t)$ and the collision rate $\gamma(t)$ are modulated in time, we recover the most general scenario for Eq.~(\ref{epsilon_two_terms}). 
	As a key insight for future analyses involving time-varying Drude model, Eq.~(\ref{epsilon_two_terms}) reveals that a time-dependent collision rate does not simply translate into an effective $\gamma(t)$ entering the denominator of the Drude expression—neither under first-order approximation, nor when higher-order effects are included. 
	Rather, the temporal dependency of $\gamma(t)$ shapes the kernel $I(t,\omega)$ itself as a non-separable function in its temporal and frequency dependencies. This non-separability arises from the time-dependent loss modulation, giving rise to the observed behavior in the simulations. As shown in the third row of \hyperref[three_models_epsilon_t_w]{Fig.~\ref*{three_models_epsilon_t_w}}, a time-dependent collision rate not only attenuates the amplitude of the oscillations observed in the previous (middle row) case, but also produces a pronounced blurring along the temporal axis, which is related with breaking the ratio of proportionality between $\omega_{p}(t)$ and $\gamma(t)$ due to a slightly different amplitude $\delta\omega_{p,2}$ in the later (see appendix ~\ref{sec:appendix}). 
This blurring arises from the accumulation of phase mismatch between modulation cycles, producing a smeared permittivity profile.

\subsection{TIME DOMAIN}\label{sec:two_times}
In the two-times formalism, the dielectric response is described by a temporally non-local relation between the polarization field $P(t)$ and the electric field $E(t)$. This response reflects the medium's memory and is expressed through a convolution over the past evolution of the field. Here, the variables are $t$, which denotes the current observation time, whereas $t'$ represents the action time at which the field is applied. Usually, another variable is conveniently defined as $\tau=t-t'$. In static media, the susceptibility depends only on this time difference $\tau$, i.e., $\chi(\tau)$, which allows a direct Fourier transform to $\chi(\omega)$ and consequently to $\varepsilon(\omega)$. However, when $\omega_p(t)$ or $\gamma(t)$ are explicitly time-dependent, this time-translation invariance is broken, and then the susceptibility kernel depends on $t$ and $t'$, rather than solely on $\tau$. Under these conditions, the dielectric response must describe cumulative memory effects of the instantaneous temporal modulation. Starting from the mixed Laplace-time susceptibility kernel in Eq.~(\ref{kernel_mixed_laplace}), the inverse Laplace transform $\mathcal{L}^{-1}\!\left[e^{s(t''-t)}\right](\tau)$, with $\tau = t - t'$, yields to:

\begin{equation}
	\chi(t,t-t') =\Theta(t - t') I_{\gamma}(t')\, 
	\int_{t'}^{t} dt''\, 
	\frac{\omega_{p}^{2}(t'')}{I_{\gamma}(t'')},
	\label{kernel_two_times}
\end{equation}
where $\Theta(t - t')$ enforces causality, ensuring the medium responds only to past fields ($t \ge t'$), i.e., $\chi(t,t-t')=0$ for $t<t'$. The polarization then reads:
\begin{equation}
	P(t) = \varepsilon_{0}\int_{-\infty}^{t} dt'\,\chi(t,t-t')
	E(t')\,.
	\label{polarization_two_times}
\end{equation}
Therefore, the Eqs.~(\ref{kernel_two_times}) and (\ref{polarization_two_times}) provide a closed form  for a time-varying Drude model in a two-times domain. The susceptibility kernel~(\ref{kernel_two_times}) also explicitly incorporates a direct physical interpretation of modulation, allowing a direct numerical implementation and a clear bridge to the static limit $\chi(\tau)$. For a linear response, the displacement field is given by:
\begin{equation}
	D(t) = \varepsilon_{0}\int_{-\infty}^{t} dt'\,\varepsilon(t,t-t')
	E(t')\,,
	\label{D_two_times}
\end{equation}
where the two-times permittivity reads
\begin{equation}
	\varepsilon(t,t-t') = 
	\varepsilon_{\infty}\,\delta(t - t') +
	\chi(t,t-t').
	\label{varepsilon_two_times}
\end{equation}
Noteworthy, in contrast with the mixed time–frequency susceptibility~(\ref{kernel_mixed}) and permittivity~(\ref{epsilon_two_terms}), whose frequency components involve complex terms—mixing causality with Fourier-phase information, obscuring the physical interpretation—, their two-times counterparts, Eqs.~(\ref{kernel_two_times}) and (\ref{varepsilon_two_times}), remain purely real since all macroscopic parameters are real. In addition, causality is explicitly guaranteed by the Heaviside distribution $\Theta(t-t')$. Hence, this causal structure implies that $\chi(t,t-t')$ acts as a memory kernel. Notice that in the limits of constant $\gamma(t)$, constant $\omega_p^2(t)$, or both, Eq. (\ref{kernel_two_times}) reduces to an standard leaky integrator weighted by an exponential factor, simple rescaling, or the usual Drude kernel, respectively. A comparison between a first-order~\cite{stepanov1976dielectric}, higher-order ~\cite{horsley2025macroscopic}, and time-dependent collision rate time-varying Drude model is represented in \hyperref[three_models_chi_t_tprime]{Fig.~\ref*{three_models_chi_t_tprime}}, showing the behavior of the susceptibility kernel $\chi(t,t')$ for the same aforementioned modulation profile for plasma frequency $\omega_{p}(t)$ and collision rate $\gamma(t)$.

In \hyperref[three_models_chi_t_tprime]{Fig.~\ref*{three_models_chi_t_tprime}}, the first row illustrates the susceptibility kernel for the first-order model, where only the plasma frequency $\omega_p(t)$ varies in time while the collision rate $\gamma$ remains constant. This modulation alone breaks time-translation invariance, but the structure of the kernel still follows the critically damped Drude form in $t-t'$, simply weighted by $\omega_p^{2}(t)$. Under these approach, Eq.~(\ref{kernel_two_times}) reduces to:

\begin{equation}
	\chi(t,t-t') \approx \Theta(t-t') \frac{\omega_p^2(t)}{\gamma}\left[1-e^{-\gamma(t-t')}\right].
	\label{chi_first_order}
\end{equation}

\onecolumngrid
\begin{center}
	\begin{figure}[H]
		\centering
		\includegraphics[scale=0.94]{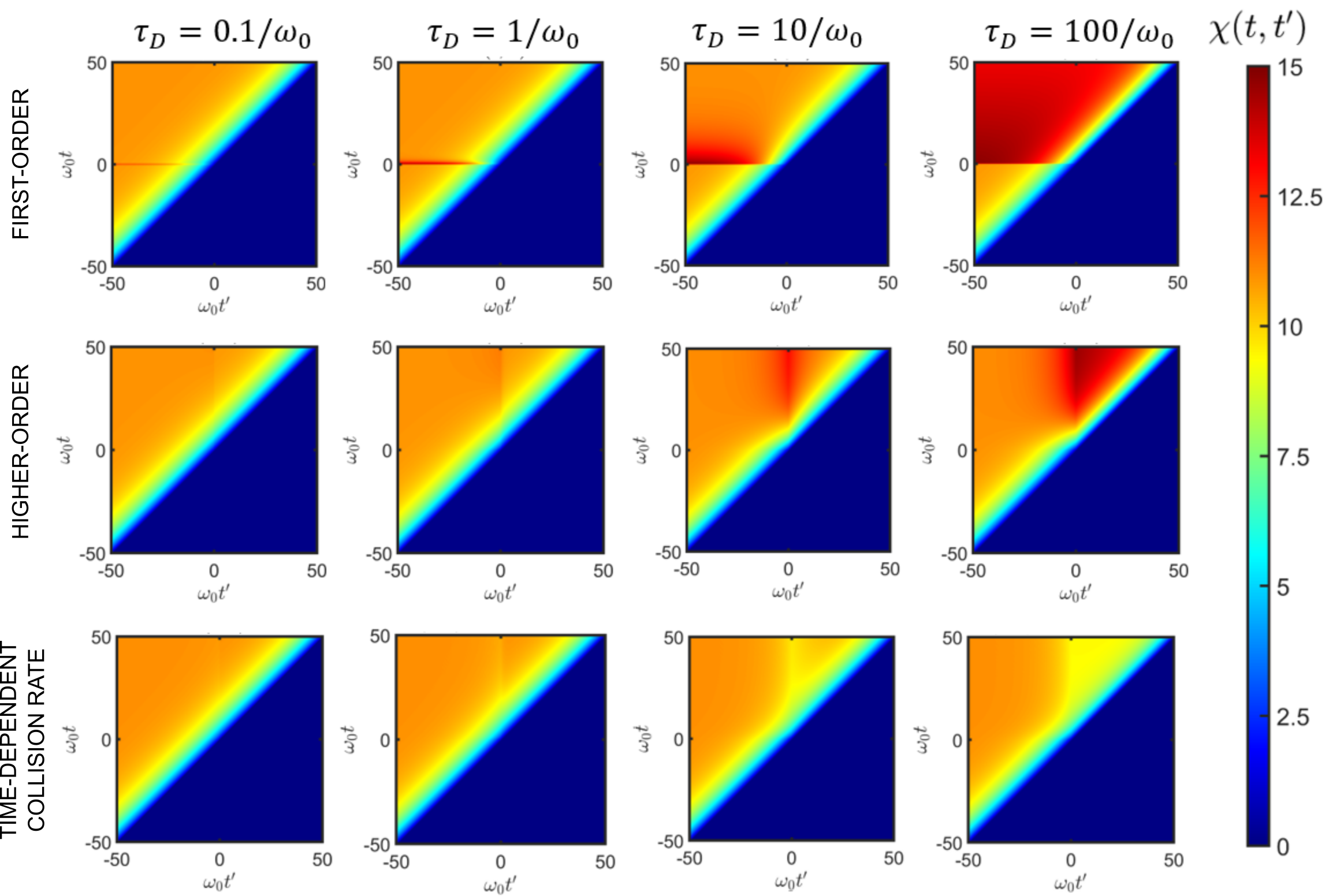}
		\caption{
			\label{three_models_chi_t_tprime}
			Susceptibility kernel $\chi(t,t')$. In all panels the plasma frequency follows $\omega_{p}^{2}(t)=\omega_0^{2}+\delta\omega_{p,1}^{2} e^{-t/\tau_L}\left[1+\tanh(t/\tau_D)\right]$,
			with $\omega_{p,\max}^{2}=1.4,\omega_{0}^{2}$ and $\omega_0\tau_L=0.1$. Top row: first-order (Stepanov) model~\cite{stepanov1976dielectric}, where only $\omega_{p}^{2}(t)$ varies in time and the collision rate is fixed at $\gamma=0.1\omega_0$. Middle row: higher-order model~\cite{horsley2025macroscopic}, which incorporates the non-instantaneous response generated by rapid variations of $\omega_{p}^{2}(t)$ while keeping $\gamma=0.1,\omega_0$ constant. Bottom row: fully time-dependent model with both parameters varying, where $\gamma(t)=0.1,\omega_p(t)$ and the modulation amplitude $\delta\omega_{p,2}^{2}$ is adjusted so that $\gamma_{\max}=0.14,\omega_0$. In this case the joint time dependence of $\omega_p(t)$ and $\gamma(t)$ reshapes the kernel across the $(t,t')$ domain and can invert the local maximum at $t'=0$ into a minimum. The four columns correspond to $\omega_0\tau_D = 0.1$, $1$, $10$, and $100$, showing the transition from ultrashort symmetric modulations to strongly asymmetric profiles dominated by the hyperbolic-tangent envelope.
		}
	\end{figure}
\clearpage        
\end{center}
\twocolumngrid
\noindent As shown in the first row of \hyperref[three_models_chi_t_tprime]{Fig.~\ref*{three_models_chi_t_tprime}}, the first-order susceptibility~(\ref{chi_first_order}) rises sharply and attains its maximum at the observation time $t=0$ for any $t' < t$. This means that when the electric field $E(t)$ drives a Drude medium described by the susceptibility kernel~(\ref{chi_first_order}), the resulting polarization field~(\ref{polarization_two_times}) will be weighted by the onset of the instantaneous modulation in $\omega_p^{2}(t)$. Thus, the polarization field will display an underdamped behavior characterized by a transient overshoot before relaxing toward its steady value. For a symmetrical modulation ($\tau_L=\tau_D=0.1/\omega_0$) of the plasma frequency, the temporal profile of $\chi(t,t')$ remains highly symmetric and narrowly confined, so the resulting susceptibility and polarization field remain essentially almost indistinguishable from their static Drude counterparts. As $\tau_D$ increases for a given $\tau_L$, the modulation of $\omega_p^2(t)$ introduces a slight stretching of the kernel along the $t$ axis. This effect becomes most evident for highly asymmetric modulations ($\tau_D = 100/\omega_0$).

However, as the instantaneous modulation of $\omega_p^2(t)$ becomes faster, the adiabatic assumption underlying the first-order model is no longer justified under its own terms; the chosen value of $\tau_L$ was selected precisely to expose this limitation of the first-order model. Thus, the susceptibility cannot be approximated by a Drude kernel merely weighted by $\omega_p(t)$. Instead, the instantaneous value of $\omega_p^{2}(t)$ must be integrated over the interval $(t',t)$, as required by Eq.~(\ref{kernel_two_times}). For constant $\gamma$, the higher-order susceptibility takes the form:

\begin{equation}
	\chi(t,t-t') = \Theta(t-t')\, e^{\gamma t'} \int_{t'}^{t} dt'' \,\omega_p^2(t'') \, e^{-\gamma t''}.
	\label{chi_higher_order}
\end{equation}

In this limit case, the modulation of $\omega_p^2(t)$ is weighted by the exponential memory factor associated with losses. Crucially, because the kernel integrates $\omega_p^2(t'')$ over $(t',t)$, the area under the curve in Eq.~(\ref{chi_higher_order}) becomes a key factor. This means that the pulsed modulation enhances the excitation of the polarization field over a given interval, performing a sort of temporal gating. For an asymmetric pulse of fixed amplitude $\delta\omega_{p,1}$, rapid modulations contribute little within the integration window, whereas slower decays produce a larger area, partially tempered by the exponential factor $\gamma^{-1}$. As shown in the second row of \hyperref[three_models_chi_t_tprime]{Fig.~\ref*{three_models_chi_t_tprime}}, the susceptibility now peaks at $t'=0$, i.e., when the action time matches the onset of the modulation. As a noteworthy signature of this higher-order model, here the approach to this maximum is smooth, in contrast with the abrupt behavior of the first-order model. This reflects the fact that the polarization field incorporates the loading time $\tau_L$ through the integral in~(\ref{chi_higher_order}). While ultrashort modulations do not enhance the susceptibility or the resulting polarization field, slow decay times $\tau_D$ broaden the kernel in both variables and yield a stronger polarization response, i.e., a higher steady state at later observation times.

Finally, the third row in \hyperref[three_models_chi_t_tprime]{Fig.~\ref*{three_models_chi_t_tprime}} illustrates a more general case, where both $\omega_p(t)$ and $\gamma(t)$ vary in time. In this regime, the coupled evolution of $\omega_p^{2}(t'')$ and the integrating factor associated with $\gamma(t'')$ affects the entire integration domain in Eq.~(\ref{kernel_two_times}), preventing any factorization of the kernel in the $(t,t')$ plane. Because the time-dependent collision rate governs both the decay of the memory kernel and the redistribution of weight among earlier field contributions, the susceptibility develops a nontrivial topology that no longer resembles a Drude-like profile. This means that the pulsed modulation can actually inhibit the excitation of the polarization field when losses are strongly increased. For sufficiently large amplitudes $\delta\omega_{p,2}$, this coupling induce temporal anti-gating effects, driving the kernel’s local maximum at $t'=0$ into a minimum (see Appendix~\ref{sec:appendix}). This inversion then transfers to the polarization $P(t)$, which may exhibit a transient suppression whose duration increases with the decay time $\tau_D$.

In light of these results, the modulation of losses gives rise to distinct dynamical regimes for the susceptibility kernel~(\ref{kernel_two_times})—ranging from simple attenuation in the excitation of the polarization field, to a Drude-like behavior, and ultimately to kernel inversion when damping becomes sufficiently strong, which is the case shown in Fig.~\ref{three_models_chi_t_tprime}. These three different regimes are further detailed in the Appendix~\ref{sec:appendix}. Importantly, these behaviors acquire a clear and directly observable physical interpretation only within the two-times representation. In this framework, the balance between $\omega_p^2(t)$ and $I_{\gamma}(t)$ determines whether the polarization field is enhanced or inhibited. As a result, a time-dependent collision rate $\gamma(t)$ allows the pulsed modulation of $\omega_{p}(t)$ to selectively gate the excitation of the polarization field over a defined temporal interval, while strong losses may suppress the kernel near the activation time $t'=0$ or even induce a transient inversion. Therefore, a time-varying $\gamma(t)$ does far more than attenuate the susceptibility—it qualitatively alters its dynamics.
\subsection{FREQUENCY DOMAIN}\label{sec:two_frequencies}

In dispersive time-varying media, a monochromatic electric field $E(\omega)$ does not produce a single-frequency polarization governed by $P(\omega)=\chi(\omega)E(\omega)$. Instead, temporal diffraction redistributes energy across multiple frequencies. This motivates a two-frequency description in which the input and output frequencies, $\omega'$ and $\omega$, respectively, provide a complete spectral characterization of the medium’s response. Starting from polarization field in Eq.~(\ref{polarization_mixed_laplace}), its form in Laplace domain now will depend on the difference between output variables $s$ and input (integrated) variable $s'$:
\begin{equation}
P\left(s\right)=\varepsilon_{0}\thinspace\int_{\xi-i\infty}^{\xi+i\infty}\frac{ds'}{2\pi i}\thinspace\chi\left(s-s',s'\right)\,E\left(s'\right).
	\label{polarization_two_s}
\end{equation}
where the two Laplace susceptibility kernel is:
\begin{equation}
\chi(s-s',s')=\int_0^\infty dt \chi(t,s) e^{-(s-s')t} 
\end{equation}

Then, defining the quantity $f(t,s')=\omega_{p}^{2}(t)I(t,s')$, where the mixed time-Laplace function stands for $I(t,s)=~\int_{-\infty}^{t}dt'\thinspace\left[I_{\gamma}\left(t'\right)/I_{\gamma}\left(t\right)  \right]\thinspace e^{-s\left(t'-t\right)}$, it is possible to express the exponential terms for $\chi(t,s')$ in Eq.~(\ref{kernel_mixed_laplace}) as a total derivative. Using  the fundamental theorem of calculus allows the susceptibility kernel to be decomposed into first-order term and higher-order term, $\chi_{1}$ and $\chi_{2}$, respectively. Using shifted Laplace-transform definition with $s-s'$, the first-order term becomes
\begin{equation}
	\begin{gathered}
	\chi_{1}\left(s-s',s'\right)=\frac{1}{s'}f\left(s-s',s'\right),
	\end{gathered}
	\label{chi_1}
\end{equation}
which depends on the difference $s - s'$, and exhibits a simple pole in $s'=0$. On the other hand, convolution properties over higher-order term reads:
\begin{equation}
	\chi_{2}\left(s-s',s'\right)=-\frac{s-s'}{s's}f\left(s-s',s'\right),
	\label{chi_2}
\end{equation}
which has poles in both variables, a pole at the input variable $s' = 0$ and another at the output complex variable $s = 0$. Next, performing a Laplace-to-Fourier transform projecting  both $s$ and $s$ onto the imaginary axis, i.e., $s\mapsto-i\omega + 0^{+}$ and $s'\mapsto-i\omega' + 0^{+}$, the susceptibility kernel in the two-frequencies domain is written as follows:
\begin{equation}
	\begin{gathered}
\chi(\omega-\omega',\omega)=\frac{f\left(\omega-\omega',\omega'\right)}{-i\omega'+0^{+}}+\frac{i\left(\omega-\omega'\right)f\left(\omega-\omega',\omega'\right)}{\left(-i\omega'+0^{+}\right)\left(-i\omega+0^{+}\right)},
	\end{gathered}
	\label{kernel_two_omegas}
\end{equation}
where $f\left(\omega-\omega',\omega'\right)$ is the shifted Fourier transform of  $f(t,\omega')=\omega_{p}^{2}(t)I(t,\omega')$. Noticing $\omega' \mapsto \omega' + i0^{+}$, the polarization field then reads
\begin{equation}
P\left(\omega\right)=\varepsilon_{0}\thinspace\int_{-\infty}^{+\infty}\frac{d\omega'}{2\pi}\thinspace\chi\left(\omega-\omega',\omega'\right)\,\tilde{E}\left(\omega'\right).
\label{P_omega}
\end{equation}

Considering a linear media, the displacement field in the frequency domain is given by:
\begin{equation}
D\left(\omega\right)=\varepsilon_{0}\int_{-\infty}^{+\infty}\frac{d\omega'}{2\pi}\thinspace\varepsilon\left(\omega-\omega',\omega'\right)\,\tilde{E}\left(\omega'\right)
	\label{D_omega}
\end{equation}
with the two-frequencies permittivity:
\begin{equation}
	\varepsilon\left(\omega-\omega',\omega'\right)=2\pi\varepsilon_{\infty}\delta(\omega-\omega')+\chi\left(\omega-\omega',\omega'\right).
	\label{varepsilon_omega}
\end{equation}

Combining Eqs.~(\ref{chi_1}), (\ref{chi_2}), and (\ref{varepsilon_omega}), the permittivity, once projected onto two frequencies, reads:

\begin{widetext}
	\begin{equation}
		\varepsilon\left(\omega-\omega',\omega'\right)=2\pi\varepsilon_{\infty}\delta\left(\omega-\omega'\right)+\frac{f\left(\omega-\omega',\omega'\right)}{-i\omega'+0^{+}}+\frac{i\left(\omega-\omega'\right)f\left(\omega-\omega',\omega'\right)}{\left(-i\omega'+0^{+}\right)\left(-i\omega+0^{+}\right)}.
		\label{epsilon_two_omegas}
	\end{equation}
\end{widetext}

In this two-frequencies representation, Eq.~(\ref{epsilon_two_omegas}) defines the kernel $\varepsilon\left(\omega-\omega',\omega'\right)$, which maps an excitation $\omega'$ to a response $\omega$, through a static diagonal term, a modulation-induced mixing term, and higher-order contribution stemming from temporal derivatives.

\begin{center}
	\begin{figure}[H]
		\centering
		\includegraphics[scale=0.90]{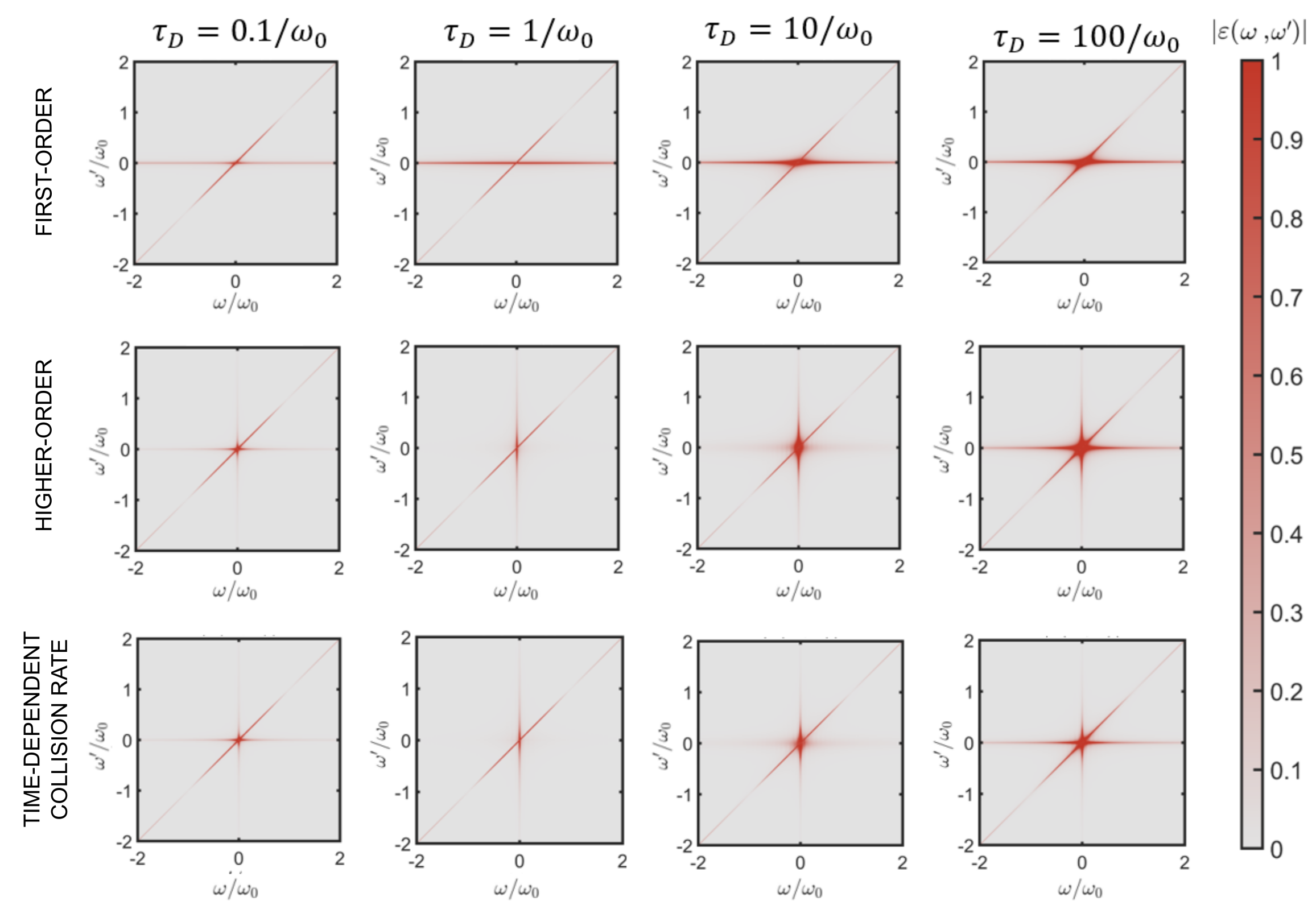}
		\caption{
			\label{epsilon_w_wprime}
			Absolute value of permittivity $\varepsilon(\omega,\omega')$. The plasma frequency follows $\omega_{p}^{2}(t)=\omega_0^{2}+\delta\omega_{p,1}^{2} e^{-t/\tau_L}\left[1+\tanh(t/\tau_D)\right]$,
			with $\omega_{p,\max}^{2}=1.4,\omega_{0}^{2}$ and $\omega_0\tau_L=0.1$. Top row: first-order (Stepanov) model~\cite{stepanov1976dielectric}, with time-varying $\omega_p(t)$ and fixed collision rate $\gamma=0.1\omega_0$, showing near-diagonal coupling $\omega=\omega'$ that broadens as $\tau_D$ increases. Middle row: higher-order model~\cite{horsley2025macroscopic}, including non-instantaneous effects of rapid $\omega_p(t)$ variations while keeping $\gamma=0.1\omega_0$ constant. The contribution of higher-order derivatives generates a low-frequency feature, redistributing spectral weight into near-diagonal and output low-frequency channels. As $\tau_D$ increases, this feature diminishes and partially cancels the quasi-static response, before reappearing for longer modulation times. Bottom row: fully time-dependent model, with both $\omega_p(t)$ and $\gamma(t)$ varying; the kernel is non-separable and primarily attenuated across all frequencies, without new qualitative features. The four columns correspond to $\omega_0\tau_D = 0.1$, $1$, $10$, and $100$, showing the transition from ultrashort symmetric modulations to asymmetric profiles dominated by the hyperbolic-tangent envelope.
		}
	\end{figure}
\end{center}

To assess Eq.~(\ref{epsilon_two_omegas}), we compute the FFT of Eq.~(\ref{epsilon_two_terms}) and compare the two-frequency permittivity kernels for the first-order \cite{stepanov1976dielectric}, higher-order, and time-dependent collision-rate Drude models. The same profile used throughout this work for $\omega_{p}^2(t)$ and $\gamma(t)$ is revisited in \hyperref[epsilon_w_wprime]{Fig.~\ref*{epsilon_w_wprime}}. We note, however, that constructing $\varepsilon(\omega,\omega')$ via FFT of $\varepsilon(t,\omega)$ introduces numerical artifacts absent in the analytical formulation: numerical broadening of the diagonal $\delta(\omega-\omega')$ terms, causal principal-value and $0^{+}$ prescriptions are not faithfully reproduced, singular behavior near $\omega=0$, among others.

In \hyperref[epsilon_w_wprime]{Fig.~\ref*{epsilon_w_wprime}}, the first row shows the first-order model permittivity, where the last term of Eq.~(\ref{epsilon_two_omegas}) is neglected, the plasma frequency $\omega_p(t)$ varies in time while collision rate $\gamma$ remains constant. Hence, Eq.~(\ref{epsilon_two_omegas}) reads:
\begin{equation}
	\varepsilon(\omega-\omega',\omega') =
	2\pi\varepsilon_\infty \delta(\omega-\omega')
	-\frac{\omega_p^{2}(\omega-\omega',\omega')}{\omega'(\omega'+i\gamma)} ,
	\label{epsilon_first_order}
\end{equation}
A time-varying $\omega_p(t)$ alone couples nearby frequencies via non-zero Fourier components ${\omega}_p^{2}(\omega-\omega',\omega')$ away from the diagonal $\omega=\omega'$. The Drude denominator retains its low-frequency divergence, while the numerator governs energy transfer from $\omega'$ to neighboring $\omega$. When $\omega'=\omega_0$, the diagonal of $|\varepsilon(\omega,\omega')|$ shows a minimum associated with the ENZ response. Symmetric pulses ($\tau_L=\tau_D=0.1/\omega_0$) yield a narrow diagonal kernel, while increasing $\tau_D$ slightly broadens the coupling for highly asymmetric modulations ($\tau_D=100/\omega_0$), since fixing the peak of $\omega_p^2(t)$ increases the modulation area—contrary to the expected if Fourier transforming equal-area pulses.

The second row in \hyperref[epsilon_w_wprime]{Fig.~\ref*{epsilon_w_wprime}} shows the two-frequency permittivity kernel $|\varepsilon(\omega,\omega')|$ for the higher-order model, in which the third term accounting the fast modulation of $\omega_p^2(t)$ is included, while $\gamma$ remains constant. The permittivity is given then by:

\begin{equation}
	\begin{gathered}
		\varepsilon\left(\omega-\omega',\omega'\right)
		= 2\pi\varepsilon_{\infty}\delta(\omega-\omega')
		- \frac{\omega^2_p(\omega-\omega',\omega')}{\omega'(\omega'+i\gamma)} \\[1pt]
		- i\,\frac{\omega-\omega'}{\omega'\omega}
		\frac{\omega^2_p(\omega-\omega',\omega')}{\omega'(\omega'+i\gamma)}
	\end{gathered}
	\label{epsilon_higher_order}
\end{equation}

The first two terms in Eq.~(\ref{epsilon_higher_order}) coincide with the first-order model, while the third term captures non-adiabatic effects from rapid plasma-frequency variations and introduces a pole at $\omega=0$. When the input frequency is close to the ENZ value $\omega_0$, the main diagonal of $|{\varepsilon}(\omega,\omega')|$ again exhibits the characteristic suppression of the Drude response at the ENZ point. Regarding the effects of modulation times, when considering ultrashort pulses ($\tau_D=0.1/\omega_0$), the kernel keeps the diagonal feature. The quasi-static feature near $\omega' \approx 0$ is still present, while the new low-frequency output component remains confined to the bandwidth of $\omega_p^2(t)$; since the ultra-fast excitation is short-lived, its influence across a broader frequency range is limited. As $\tau_D$ increases to mid values ($\tau_D \sim 1/\omega_0$), the low-frequency contribution around $\omega'~=~0$ diminishes. For longer modulation times ($\tau_D = 10/\omega_0$ and $100/\omega_0$), the low-frequency line for input frequencies reappears.

Finally, the third row in \hyperref[epsilon_w_wprime]{Fig.~\ref*{epsilon_w_wprime}} shows the two-frequency permittivity kernel for the general case where both $\omega_p(t)$ and $\gamma(t)$ vary in time. In this regime, the coupled dynamics of these parameters produce a non-separable kernel that depends on the past history of the collision rate, so that the second term in Eq.~(\ref{epsilon_two_omegas}) no longer corresponds to the Fourier transform of an instantaneous modulation. As a result, the overall spectral amplitude is reduced compared with the higher-order model, reflecting the less efficient redistribution of energy across frequencies when losses are time modulated, while the qualitative spectral structure remains unchanged.

\onecolumngrid
\twocolumngrid
\subsubsection{Reflection in a thin time-varying slab}

To understand how a finite material region modifies the spectral components of a field, it is useful to consider the interplay of spatial propagation and frequency mixing. In what follows, the magnitude and structure of the reflection matrices (for further details of those numerical tools, see Appendix~\ref{sec:numerical_tools}) are examined for a thin slab of thickness $d = 300\,\mathrm{nm}$. \hyperref[r_slab]{Fig.~\ref*{r_slab}} summarizes these behaviors by displaying the order of magnitude of $|r_{\mathrm{slab}}(\omega,\omega')|$ for three models across several modulation times: (i) the first-order model with time-dependent $\omega_p(t)$ and fixed $\gamma$ in Eq.~(\ref{epsilon_first_order}), (ii) the higher-order model accounting for non-adiabatic modulation effects in Eq.~(\ref{epsilon_higher_order}), and (iii) the fully time-dependent model in which both $\omega_p(t)$ and $\gamma(t)$ vary according to Eq.~(\ref{epsilon_two_omegas}).


\onecolumngrid
\begin{center}
	\begin{figure}[H]
		\centering
		\includegraphics[scale=0.92]{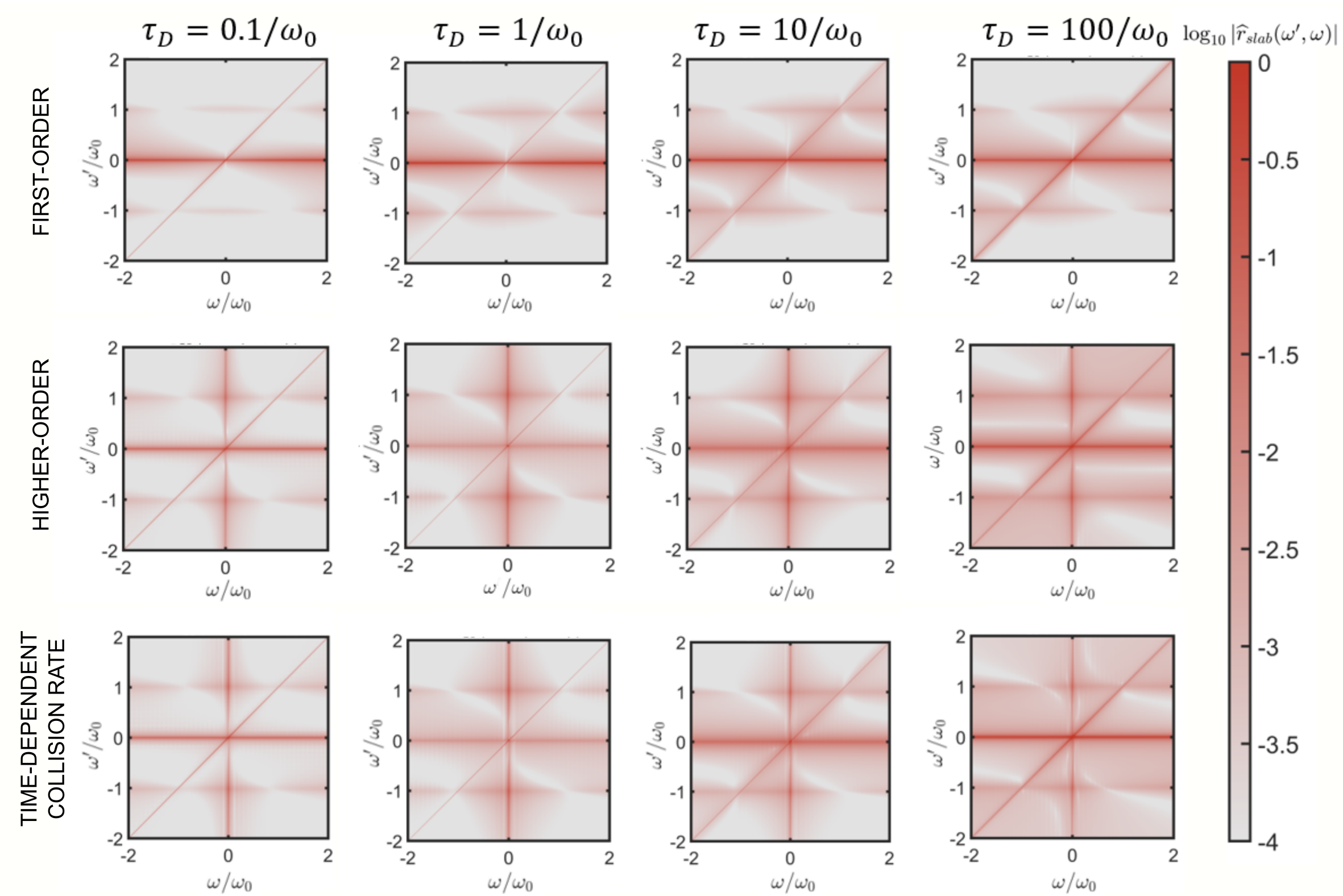}
		\caption{
			\label{r_slab}
			Magnitude of the reflection operator $|r_{\mathrm{slab}}(\omega,\omega')|$ for a thin slab of thickness $d=300\,\mathrm{nm}$. Red indicates strong reflection, while white denotes weak coupling between input frequency $\omega'$ (vertical axis) and output frequency $\omega$ (horizontal axis). The plasma frequency evolves as $\omega_p^2(t)=\omega_0^{2}+\delta\omega_{p,1}^{2}e^{-t/\tau_L}[1+\tanh(t/\tau_D)]$, with $\omega_{p,\max}^2=1.4\omega_0^2$ and $\omega_0\tau_L=0.1$. Top row: First-order (Stepanov) model~\cite{stepanov1976dielectric} with time-varying $\omega_p(t)$ and fixed $\gamma=0.1\omega_0$, showing predominantly diagonal coupling ($\omega \approx \omega'$) that broadens as $\tau_D$ increases. Middle row: Higher-order model~\cite{horsley2025macroscopic}, which incorporates non-instantaneous modulation effects. The contribution of higher-order derivatives generates a low-frequency feature that redistributes spectral weight between near-diagonal and $\omega'\!\approx 0$ channels; this structure weakens for intermediate $\tau_D$ and re-emerges for longer modulations. Bottom row: Fully time-dependent model, with both $\omega_p(t)$ and $\gamma(t)$ varying, producing a non-separable kernel that is more strongly attenuated across frequencies and does not introduce additional qualitative features.  Columns correspond to $\omega_0\tau_D = 0.1,\,1,\,10,$ and $100$, tracing the transition from ultrashort symmetric modulations to slower, asymmetric profiles dominated by the hyperbolic-tangent envelope.
		}
	\end{figure}
\clearpage
\end{center}
\twocolumngrid

In the first row of \hyperref[r_slab]{Fig.~\ref*{r_slab}}, corresponding to the first-order permittivity (\ref{epsilon_first_order}), the reflection operator shows a strong diagonal band at $\omega = \omega'$, indicating narrow, frequency-preserving reflection (i.e., elastic scattering), alongside weak off-diagonal regions arising from the modulation of $\omega_p^2(t)$. Since $\omega'=\omega_0$ corresponds to the ENZ point, reflection is enhanced and extended to frequencies below $\omega_0$. As the modulation time $\tau_D$ increases, the off-diagonal bands broaden, redistributing spectral components over a wider frequency range and promoting more symmetric frequency mixing, while the main passband progressively widens.

For the higher-order model, using the non-adiabatic permittivity (\ref{epsilon_higher_order}) introduces additional structures in the frequency spectra. Notably, as it is shown in the second row of \hyperref[r_slab]{Fig.~\ref*{r_slab}}, the higher-order terms redistributes spectral weight from the near-diagonal region into channels near $\omega \approx 0$, generating an output baseband component absent in the first-order case. Overall, the region of significant reflection widens into a cross-like domain, with arms along both the diagonal $\omega = \omega'$ and the low-frequency axes $\omega \approx 0$ and $\omega' \approx 0$. Again, as the modulation times $\tau_D$ increases, the structure reorganizes, and the baseband signature intensifies.

Finally, the third row of \hyperref[r_slab]{Fig.~\ref*{r_slab}} illustrates the fully time-dependent model (\ref{epsilon_two_omegas}), where both the plasma frequency $\omega_p(t)$ and the collision rate $\gamma(t)$ vary in time. In this regime, the reflection operator resembles that of the higher-order model, retaining cross-like features, but these are now attenuated and broadened due to the time-dependent collision rate. As the modulation time $\tau_D$ increases, the kernel evolves from a confined cross-like profile with an output pass-band into the largest effective bandwidth among the three models.

\section{CONCLUSIONS}
In the present work, we developed a generalization of the time-varying Drude model to account for all macroscopic parameters—carrier density $N(t)$, effective mass $m(t)$, and collision rate $\gamma(t)$—as explicit functions of time. Analytical expressions for the polarization, susceptibility, displacement field, and permittivity were derived across three complementary representations: (i) mixed time–frequency domain $(t,\omega)$, (ii) two-times domain $(t,t')$, and (iii) two-frequencies domain $(\omega,\omega')$. Within those, we derived expressions for three scenarios: (i) a first-order model valid for adiabatic modulations of $\omega_p(t)$ with constant $\gamma$, (ii) a higher-order model capturing rapid temporal modulations of $\omega_p(t)$ with constant $\gamma$, and (iii) a fully generalized higher-order model incorporating a time-dependent collision rate $\gamma(t)$. For illustrative comparison, the temporal modulation of the parameters was modeled using asymmetric pulse profiles.

In the mixed time–frequency domain ($t,\omega$), Laplace transform techniques projected onto the frequency axis revealed that $\gamma(t)$ not only attenuates oscillations in the higher-order model but also introduces temporal blurring, showing that losses modify the dynamics beyond simple scaling effects. In the two-times domain ($t,t'$), a time-varying collision rate enables selective temporal gating of the polarization field, and strong damping can smooth out or suppress the kernel near the action time, producing anti-gating effects. In the two-frequencies domain ($\omega,\omega'$), higher-order corrections due to non-adiabatic modulation significantly alter spectral features, whereas time-dependent $\gamma(t)$ produces uniform attenuation without qualitatively changing the spectral distribution. These behaviors were illustrated computing reflection coefficient for a thin slab, showing that joint modulation of $\omega_p(t)$ and $\gamma(t)$ yields the largest effective bandwidth, albeit with pronounced attenuation.

Overall, this work provides a unified framework for analyzing time-dependent Drude media, highlighting the role of non-adiabatic effects and time-varying losses. In the context of temporal metamaterials, where dispersion and losses play a central role \cite{vazquez2023incandescent,vertiz2025dispersion}, we expect this formalism to support future studies on anisotropic dispersive time-varying media \cite{xu2021complete,pacheco2021spatiotemporal}, non-linear effects \cite{mock2022multiorder,zubyuk2022externally,wang2022optical,tirole2024second}, or converging with microscopic descriptions of TCO nonlinearities \cite{sarkar2023electronic} into a simple macroscopic model, with possible extensions to non-linear, relativistic, or quantum systems, opening new avenues for active control of light–matter interactions in time-varying media.

\subsection{Appendix: Asymmetric pulse profile}\label{sec:figures}
Here we illustrate a given example of pulsed temporal profiles for both $\omega_{p}^2(t)$ and the collision rate $\gamma(t)$ used throughout this work.

\begin{figure}[H]
	\centering
	\includegraphics[scale=0.35]{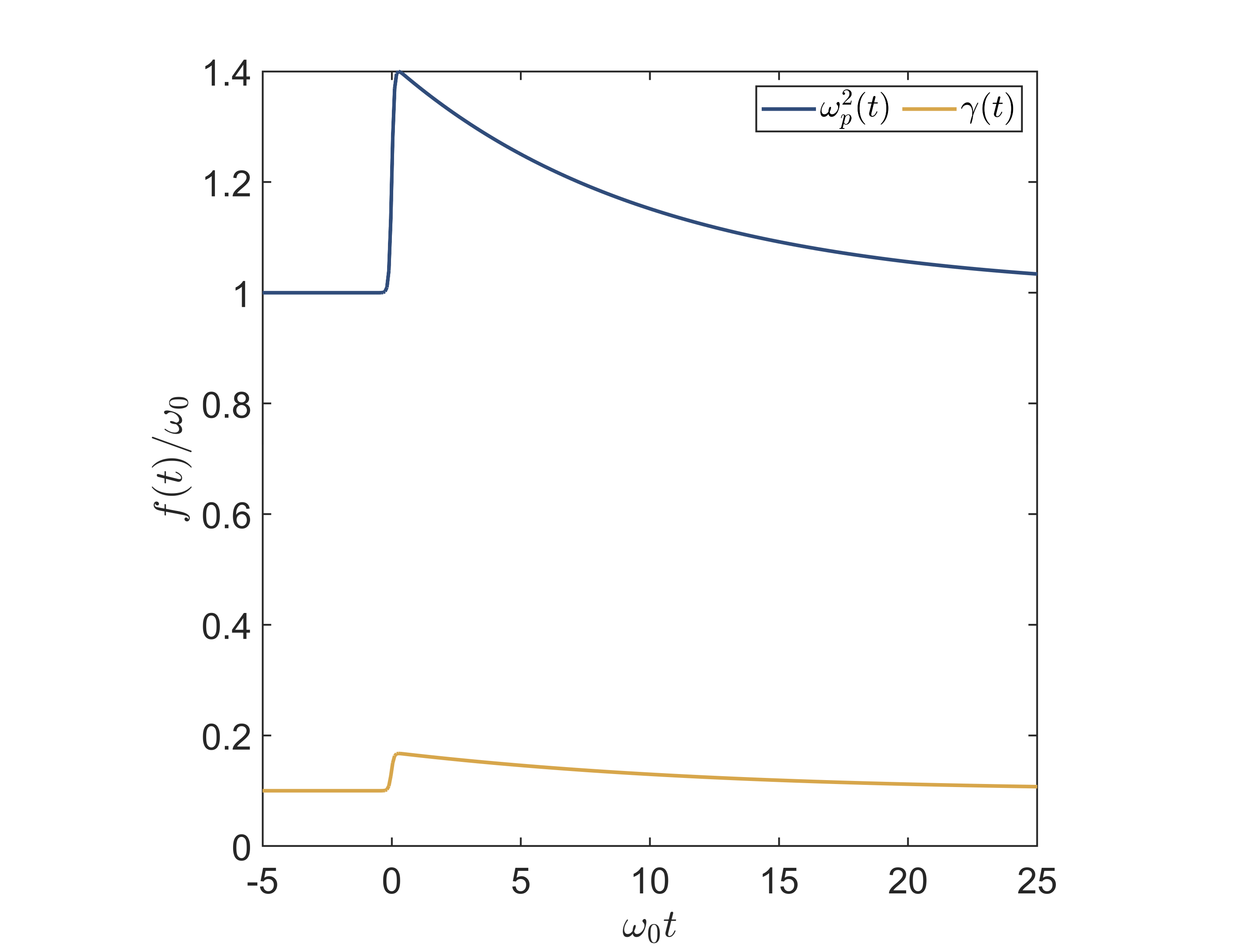}
	\caption{
		\label{appendixA}
 $\omega^2_p(t)=\omega^2_0 + \delta\omega^2_{p,1}e^{-t/\tau_L}[1+\tanh(t/\tau_D)]$, with a maximum value $\omega^2_{p,\text{max}} = 1.4\,\omega^2_0$, $\tau_L = 0.1/\omega_0$, $\tau_D = 10/\omega_0$. The collision rate evolves dynamically as $\gamma(t) = 0.1\omega_p(t)$, with a slightly adjusted modulation amplitude $\delta\omega^2_{p,2}$, such that $\gamma_{\text{max}} = 0.141\omega_{p}^{\max}$.
	}
\end{figure}

\subsection{Appendix: time-dependent losses}\label{sec:appendix}
\begin{figure}[h!]
		\centering
		\includegraphics[scale=0.65]{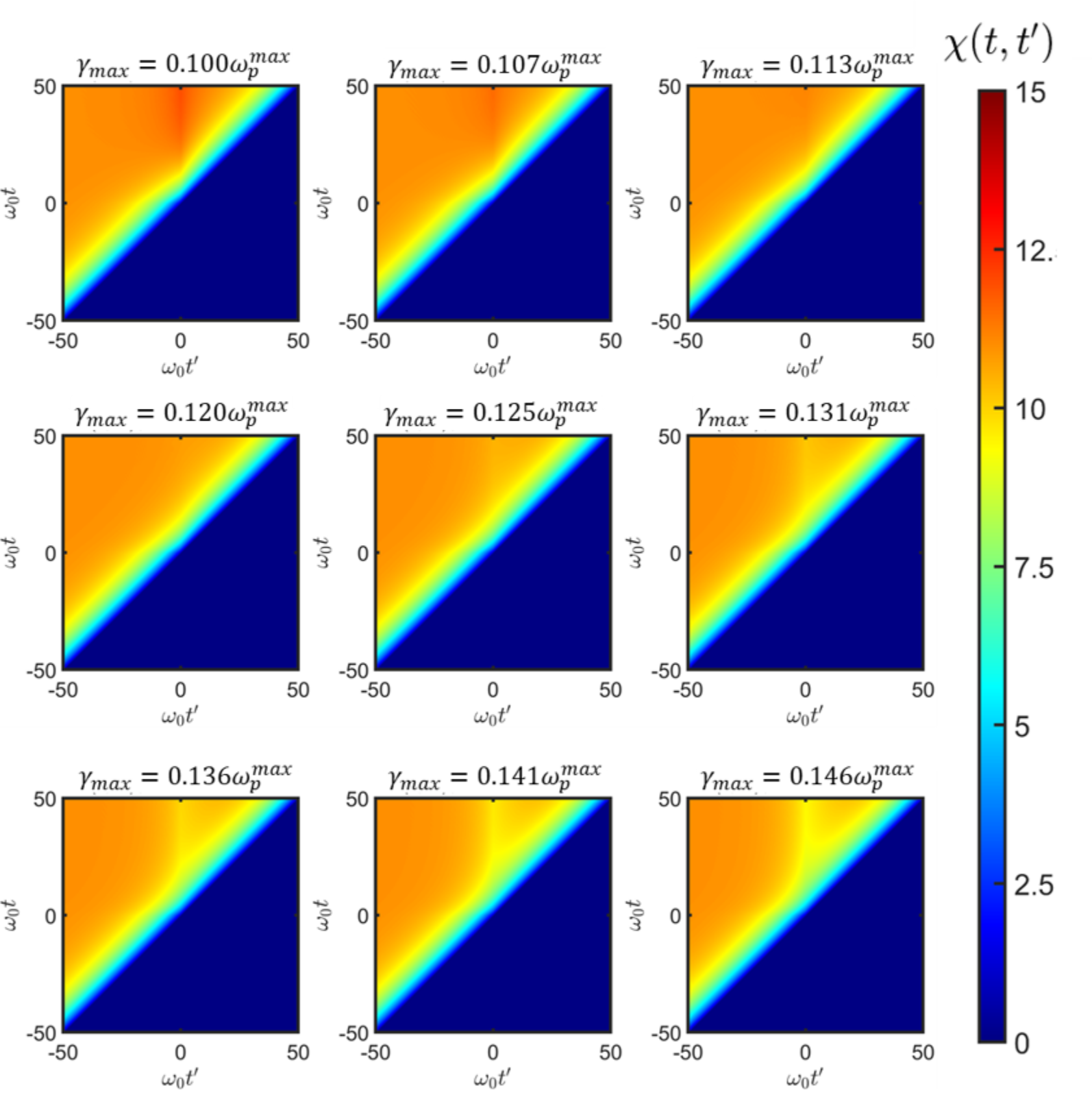}
		\caption{
			\label{chi_gamma_t}
	Susceptibility kernel $\chi(t,t')$ of the model in Eq.~(\ref{kernel_two_times}), where both $\omega_{p}^{2}(t)$ and $\gamma(t)$ are modulated by the same asymmetric pulse used throughout this work, with the damping amplitude slightly varied as indicated in the titles of each plot in \hyperref[chi_gamma_t]{Fig.~\ref*{chi_gamma_t}}. The panels illustrate three dynamical regimes induced by the modulation of $\gamma(t)$. Upper row ($\gamma=0.10\,\omega_{p}^{\max},\,0.107\,\omega_{p}^{\max},\,0.113\,\omega_{p}^{\max}$): it depicts the reference/attenuation regime, for $\gamma\approx 0.1\,\omega_{p}^{\max}$, where the kernel retains its maximum at $t'=0$ and the polarization follows the excitation with mild attenuation. Middle row ($\gamma=0.120\,\omega_{p}^{\max},\,0.125\,\omega_{p}^{\max},\,0.131\,\omega_{p}^{\max}$): it corresponds with the time-invariant/Drude-like regime, for $\gamma\gtrsim 0.1\,\omega_{p}^{\max}$, where the kernel spreads more evenly across $t$ and $t'$, producing a smoother polarization response. Bottom row ($\gamma=0.136\,\omega_{p}^{\max},\,0.141\,\omega_{p}^{\max},\,0.146\,\omega_{p}^{\max}$): it shows the inversion/high-damping regime, for $\gamma\gg 0.1\,\omega_{p}^{\max}$, where the kernel maximum at $t'=0$ flips into a minimum, leading to a transient reversal of the polarization.
}
\end{figure}

Here we give an example illustrating the behavior of the fully time-dependent susceptibility model of Eq.~(\ref{kernel_two_times}) when both $\omega_{p}^{2}(t)$ and $\gamma(t)$ are modulated by the same asymmetric pulse profile used throughout this work, with the amplitude of the damping collision rate slightly modified as indicated in the title of each plot in \hyperref[chi_gamma_t]{Fig.~\ref*{chi_gamma_t}}. The plot compares several realizations of the kernel $\chi(t,t')$ and highlights the three dynamical regimes that emerge from this modulation. 

\textbf{Gating/Low damping regime:} In this region, where $\gamma(t)\approx 0.1\,\omega_{p}(t)$, the susceptibility  $\chi(t,t')$ retains its maximum when the activation time $t'=0$, mainly subject to attenuation, as it is shown in the first row  ($\gamma=0.10\,\omega_{p}^{\max},\,0.107\,\omega_{p}^{\max},\,0.113\,\omega_{p}^{\max}$). Since the collision rate is still weak, the polarization field follows the excitation pulse almost completely. The kernel retains a clear maximum at $t'=0$, but its amplitude is slightly reduced—essentially. The system responds normally, with losses only attenuating the polarization field.
	
\textbf{Drude-like/mid damping regime:} The collision rates are stronger, and the kernel attenuates more, resembling the steady-state Drude response. Physically, the system \textit{remembers} the excitation less sharply, and the polarization builds up and decays in a smoother, more uniform way. As the collision rate increases, the susceptibility  acquires a time-invariant–like shape resembling the usual Drude kernel, illustrated in the second row ($\gamma=0.120\,\omega_{p}^{\max},\,0.125\,\omega_{p}^{\max},\,0.131$). It is worth noting that for the last element in the second row, the kernel exhibits a departure from the typical Drude-like profile, signaling that further increases in $\gamma(t)$ will lead into the inversion regime discussed below.
	
\textbf{Inversion/high-damping regime:} The collision rate dominates the dynamics, causing a redistribution of the kernel's weight such that the local maximum at $t'=0$ flips into a minimum. Physically, this means the polarization can transiently be suppressed, producing a temporary inversion of the response—a clear signature of strong anti-gating. As it is shown in the third row ($\gamma=0.136\,\omega_{p}^{\max},\,0.141\,\omega_{p}^{\max},\,0.146\,\omega_{p}^{\max}$), for sufficiently large amplitudes, the redistribution of weight induced by the integrating factor associated with the time-dependent collision rate $\gamma(t)$ dominates the dynamics of the susceptibility, driving an inversion of the local maximum into a minimum.

\subsection{Appendix: Numerical tools for reflection matrix}\label{sec:numerical_tools}
Following Ref.~\cite{horsley2023eigenpulses}, the Helmholtz equation can be written in matrix form. For an incident field along the $x$-axis, we have
\begin{equation}
	\partial_x^2 \{E,H\} + \mathbf{K}^2_{s,p} \{E,H\} = 0,
\end{equation}
where $\mathbf{K}^2_{s,p}$ governs the spatial evolution of all frequency components and depends on polarization. Assuming a TM-polarized (p) wave, the matrix operator is

\begin{equation}
	\mathbf{K}^2_p(\omega,\omega') = \frac{1}{c^2} \, \mathbf{\Omega}^2 \, \mathbf{\varepsilon}(\omega,\omega'),
	\label{Kp2}
\end{equation}
where $c$ is the speed of light and $\mathbf{\Omega}$ is a diagonal matrix of input frequencies $\omega'$, discretized to exclude $\omega' = 0$ for numerical stability. The propagation operator $\mathbf{K}_p$ is then obtained by diagonalizing $\mathbf{K}^2_p$ and taking the square root of its eigenvalues:
\begin{equation}
	\mathbf{K}_p = \mathbf{T}^{-1} \mathbf{D}^{1/2} \mathbf{T},
\end{equation}
where $\mathbf{D}$ contains the eigenvalues of $\mathbf{K}^2_p$ and $\mathbf{T}$ the corresponding eigenvectors. This construction allows all spectral components to propagate correctly through the slab, including any frequency mixing induced by time variation. The signs of the square-root eigenvalues of $\widehat K_p$ must be chosen consistently. In an $N\times N$ operator space, this choice is $2^{N}$-fold ambiguous and closely related to selecting the correct branch of the refractive index in dispersive media. For passive media, the correct branch has a positive imaginary part, ensuring decay of fields inside the medium, while in active media it can be fixed by considering the $d \to \infty$ limit of a finite slab under the assumption of stability \cite{skaar2005fresnel,pendry2000negative,cho2020digitally}. For a finite slab, the ambiguity is largely immaterial, since both signs enter symmetrically at the interfaces; nevertheless, to define the propagation operator consistently, the infinite-thickness limit is invoked, selecting the branch with eigenvalues of $\widehat K_p$ having positive imaginary part.

With $\widehat K_p$ known, the continuity conditions at the interface $x=0$ yield the generalized reflection operator. For TM polarization, this reads \cite{horsley2023eigenpulses}:
\begin{equation}
	\widehat{\mathbf r}_{p}(\omega,\omega') =
	\left[ \mathbf{I} - \widehat{\mathbf Z}_p(\omega,\omega') \right]\cdot
	\left[ \mathbf{I} + \widehat{\mathbf Z}_p(\omega,\omega') \right]^{-1},
	\label{rp}
\end{equation}
where the impedance is generalized as:
\begin{equation}
	\widehat{\mathbf Z}_p(\omega,\omega') = c\cdot \widehat{\mathbf \Omega}^{-1}\cdot\widehat{\mathbf \varepsilon}^{-1}(\omega,\omega')\cdot \widehat{\mathbf K}_p(\omega,\omega')
	\label{Zp}
\end{equation}
For a time-varying slab of thickness $d$, the reflection operator becomes
\begin{equation}
	\widehat{\mathbf r}_{\mathrm{slab}}(\omega,\omega') =
	\left[ \widehat{\mathbf A}_{-} - \widehat{\mathbf A}_{+} \widehat{\boldsymbol\Gamma} \right]\cdot
	\left[ \widehat{\mathbf A}_{+} - \widehat{\mathbf A}_{-} \widehat{\boldsymbol\Gamma} \right]^{-1},
	\label{rslab}
\end{equation}
where it has been used that $\widehat{\mathbf A}_{\pm}=\mathbf{I}\pm\widehat{\mathbf Z}_{p}(\omega,\omega')$, and $\widehat{\mathbf \Gamma}=\exp(i\widehat{\mathbf K}_{p}d)\widehat{\mathbf A}_{-}\widehat{\mathbf A}_{+}^{-1}\exp(i\widehat{\mathbf K}_{p}d)$ \cite{horsley2023eigenpulses}.

Finally, Eqns. (\ref{Kp2})–(\ref{rslab}) provide a complete mapping between the incident and reflected spectra in a slab obeying a time-varying Drude model. In the static limit, $\widehat{\mathbf r}_{\mathrm{slab}}$ reduces to the standard Fresnel expression, while in the time-varying case generalizes this familiar result by promoting every quantity to an operator that mixes frequencies. The structure of reflection operator (\ref{rslab}) closely mirrors the ordinary multilayer formula, where the matrices $\widehat{\mathbf A}_{\pm}$ stand for the boundary conditions at two interfaces, while the operator $\widehat{\boldsymbol\Gamma}$ plays the role of a generalized phase factor. Because $\widehat{\mathbf K}_{p}$ is non-diagonal, this operator couples different frequencies, meaning that a component generated at an internal interface can re-enter the first interface at a shifted frequency. Thus, the reflection operator (\ref{rslab}) provides a compact way to account for frequency diffraction, internal interference, and impedance mismatch for time-varying medium. 
\section*{ACKNOWLEDGMENTS}
The authors acknowledge support from Grant No. PID2022-137845NB-C21 funded by MICIU/AEI/10.13039/501100011033 and FEDER, EU. J.E.V.-L. acknowledges support from project PJUPNA2025-11905.	
	\bibliography{apssamp}
\end{document}